\documentclass[journal]{IEEEtran}
\usepackage[T1]{fontenc}
\usepackage{graphicx}
\usepackage{cite}
\usepackage{float}
\usepackage{algorithm}
\usepackage{amsmath}  
\usepackage{booktabs} 
\usepackage{algorithm}
\usepackage{listings}
\usepackage{xcolor}
\usepackage{amsmath, amssymb}
\usepackage{algpseudocode}

\begin{document}
%
\title{Visual and Text Prompt Segmentation: A Novel Multi-Model Framework for Remote Sensing}

\author{Xing~Zi, 
        Kairui~Jin, 
        Xian~Tao, 
        Jun~Li, 
        Ali~Braytee, 
        Rajiv~Ratn~Shah, 
        and~Mukesh~Prasad
\thanks{Xing Zi, Jun Li, Ali Braytee, and Mukesh Prasad are with the University of Technology Sydney, Australia (e-mail: \{Xing.Zi-1, Jun.Li, Ali.Braytee, Mukesh.Prasad\}@uts.edu.au).}
\thanks{Kairui Jin is with the University of Technology Sydney, Australia (e-mail: Kairui.Jin@student.uts.edu.au).}%
\thanks{Xian Tao is with the Chinese Academy of Sciences (e-mail: Taoxian2013@ia.ac.cn).}%
\thanks{Rajiv Ratn Shah is with the Indraprastha Institute of Information Technology, Delhi (e-mail: rajivratn@iiitd.ac.in).}%
}

\maketitle

\begin{abstract}
Pixel-level segmentation is essential in remote sensing, where foundational vision models like CLIP and Segment Anything Model(SAM) have demonstrated significant capabilities in zero-shot segmentation tasks. Despite their advances, challenges specific to remote sensing remain substantial. Firstly, The SAM without clear prompt constraints, often generates redundant masks, and making post-processing more complex. Secondly, the CLIP model, mainly designed for global feature alignment in foundational models, often overlooks local objects crucial to remote sensing. This oversight leads to inaccurate recognition or misplaced focus in multi-target remote sensing imagery. Thirdly, both models have not been pre-trained on multi-scale aerial views, increasing the likelihood of detection failures. To tackle these challenges, we introduce the innovative \textbf{VTPSeg} pipeline, utilizing the strengths of Grounding DINO, CLIP, and SAM for enhanced open-vocabulary image segmentation. The \textbf{Grounding DINO+(GD+)} module generates initial candidate bounding boxes, while the \textbf{CLIP Filter++(CLIP++)} module uses a combination of visual and textual prompts to refine and filter out irrelevant object bounding boxes, ensuring that only pertinent objects are considered. Subsequently, these refined bounding boxes serve as specific prompts for the \textbf{FastSAM} model, which executes precise segmentation. Our VTPSeg is validated by experimental and ablation study results on five popular remote sensing image segmentation datasets.
\end{abstract}


%
\IEEEpeerreviewmaketitle

\section{Introduction}
Remote sensing technology captures images of the Earth's surface through airborne and satellite sensors. This technology plays a crucial role in environmental monitoring, disaster response, and urban planning. Compared to traditional ground-based survey methods, remote sensing offers broader coverage and faster data acquisition, especially in hard-to-reach or hazardous areas \cite{yuan2021review}. Remote sensing image semantic segmentation is a key task that enhances the understanding and analysis of remote sensing content by finely dividing the image into areas representing different semantic categories. Despite the rich geographical and environmental information provided by remote sensing images, their analysis faces multiple challenges, such as the need for large-scale processing and data variability caused by diverse environmental conditions \cite{Clip-LSTM}.

Traditional semantic segmentation methods are often pretrained on large datasets like ImageNet and then transferred to downstream tasks. However, zero-shot remote sensing segmentation aims to accurately locate and segment objects in remote sensing images without relying on specific category training samples \cite{zhang2023text2seg}. The emergence of foundational visual models such as CLIP \cite{radford2021learning}, Grounding Dino \cite{liu2023grounding}, and Segment Anything \cite{kirillov2023segment} has inspired new approaches in zero-shot remote sensing semantic segmentation, but there are notable deficiencies:

\emph{(1) Attention Drift.} The CLIP model primarily aligns global features and neglects local objects, leading to inaccurate recognition in images with densely distributed targets or extensive scale variations. For example, while RS-CLIP \cite{li2023rs} and RemoteCLIP \cite{liu2024remoteclip} can complete aerial view classification tasks, but they rely heavily on global features(Fig. 1a). \emph{(2) Mask Redundancy and Error.} The original SAM model tends to produce numerous redundant masks without specific prompts. Variants like Semantic Segment Anything (SSA), based on BLIP \cite{chen2023semantic}, generate many candidate masks but cannot perform specific vocabulary segmentation. \emph{(3) Multi-scale Adaptation.} As shown in Fig. 1b, models like FastSAM \cite{zhao2023fast} use text, point, and bounding box prompts to guide segmentation, but the global consistency of the CLIP model makes precise local classification difficult. Simple prompts often fail to accurately identify multiple objects in remote sensing images, especially those with large scale variations.

These challenges highlight the need to improve the accuracy of existing remote sensing VLM frameworks, which motivated us to propose the VTPSeg pipeline (Fig. 1c). To handle the high resolution and large-scale range of remote sensing images, we use \textbf{Grounding DINO+ (GD+)} as the core model for multi-scale detection. This module has three subtasks. First, it scales the image at different ratios for detailed, multi-scale object recognition. Second, it uses broader synonym candidates (e.g., "roof," "house," or "building" for "building") to reduce missed detections. Third, it removes redundant and overly large candidate boxes using non-maximum suppression (NMS). These subtasks form a pipeline for target recognition.

To address classification redundancy caused by CLIP's attention drift, we created an updated classification module, \textbf{CLIP Filter++}. We add task-irrelevant but remote sensing-related prompts to mitigate classification bias. Additionally, we use visual prompts and patch enhancement (highlighting targets with red circles) to guide CLIP's attention for fine-grained classification. The combined use of GD+ and CLIP Filter++ significantly improves performance in remote sensing object recognition.

For precise fine-grained segmentation, we use point prompts with the \textbf{FastSAM} model. By setting point prompts at the center of detection boxes, the SAM model segments objects close to these points, reducing mask redundancy and false masks.

In summary, our main contributions are as follows:

\begin{itemize}
    \item We propose training-free remote sensing image segmentation pipeline, VTPSeg, which combines multi-scale adaptation and synonym prompts to improve object recognition accuracy and efficiency.
    \item We highlight the importance of VLM models' open vocabulary recognition capability and global feature consistency, leading to the development of CLIP Filter++.
    \item We address the issue of redundant and false masks generated by SAM model variants. To solve this, we use the output of GD+ module and CLIP++ with point prompts.
    \item VTPSeg outperforms other SAM-based models in terms of MIoU and Pixel Accuracy in evaluations on five remote sensing datasets. Our method achieves nearly 10\% higher MIoU and surpasses some traditional supervised learning models.
\end{itemize}

\begin{figure}[H]
\centering
\includegraphics[width=\columnwidth]{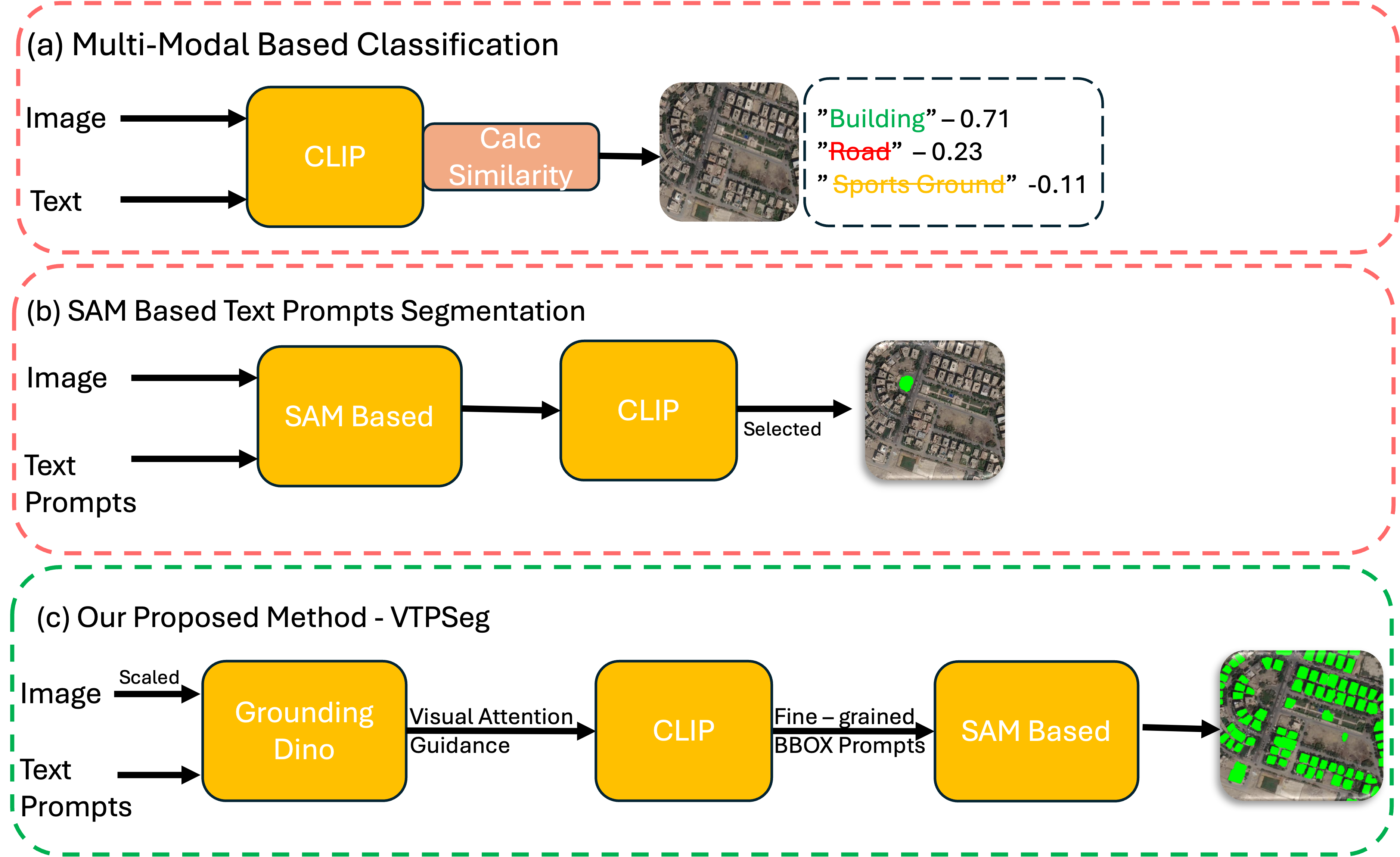}
\caption{This is remote sensing target image recognition for three different zero-shots approaches. (a) Multi-Modal Based Classification. (b) SAM Based Text Prompts Segmentation. (c) Our Proposed Method - VTPSeg}
\end{figure}

\section{Motivation}
In this paper, we empirically investigate several preliminary studies to highlight the challenges and limitations faced by current foundational vision models in the field of remote sensing. The specific research focuses are as follows:
\begin{enumerate}
    \item We investigate the number of redundant masks generated by the SAM model in the absence of specific prompt constraints and its impact on post-processing.
    \item We explore the issue of the CLIP model neglecting local objects due to global feature alignment when used as a filtering component.
    \item We further explore the impact on recognition accuracy when the CLIP model loses its open vocabulary capability after domain fine-tuning.
    \item We assess whether the introduction of multi-scale adaptation and synonym prompts can effectively reduce missed and false detections, thereby improving the accuracy and efficiency of object recognition.
\end{enumerate}
\section{Related Work}
\subsection{Remote sensing semantic segmentation}

Early techniques in remote sensing prioritized the detection of terrestrial features using predefined morphological filters, critical for revealing structural details across scales \cite{dalla2010morphological}. These methods employed differential attribute profiles to differentiate entities like roads and buildings and utilized Morphological Attribute Profiles (MAPs) to distinguish land cover based on attributes such as brightness and texture. Advancements in machine learning, particularly through deep learning, shifted focus towards using Convolutional Neural Networks (CNNs) for tasks like building damage detection in satellite imagery. This approach, which moves away from manual or rule-based methods, capitalizes on CNNs' robust feature extraction capabilities to automate and enhance the precision of identifying damaged structures \cite{xu2019building}. In improving the resolution of remote sensing images, neural networks for image super-resolution (SR) have become pivotal. Gu’s study integrates deep residual networks with Squeeze and Excitation (SE) mechanisms to boost SR performance in remote sensing \cite{gu2019deep}. The Efficient Sub-Pixel Convolutional Network (ESPCN) refines this further by minimizing convolutional layers to reduce computational demands while enabling direct feature extraction from raw images. Li's research explores an auto-encoder within the unsupervised learning framework to tackle small dataset challenges. This method involves substituting feature weights with those from VGG16 within the SSD architecture, effectively setting the stage for a pre-training regimen \cite{simonyan2014very, liu2016ssd}. These developments highlight the field's evolution from basic filter-based methods to sophisticated neural architectures, enhancing both the effectiveness and efficiency of remote sensing image analysis.

\subsection{Fundamental Models}

CNNs struggle with long-range dependencies in complex semantic segmentation tasks. Transformers, offering global semantic extraction without losing resolution, present a viable alternative to the encoder portion of traditional Fully Convolutional Networks (FCNs) \cite{zheng2021rethinking}. However, the quadratic complexity growth of traditional transformers with image size limits their use in high-resolution visual tasks. The Swin Transformer overcomes this by limiting attention to local windows and interlacing these windows between layers to capture cross-window information with lower computational demands \cite{liu2021swin}. The DINO model leverages self-supervised learning and knowledge distillation to effectively use large volumes of unlabeled data, minimizing the need for extensive annotations. It adapts well to various input sizes due to its multi-scale visual input capability. Further advancements include the DINOv2, which optimizes knowledge distillation techniques to enhance feature learning \cite{oquab2024dinov2}. The SAM model introduces cost-effective self-supervised learning strategies, using unlabeled data to achieve high segmentation accuracy \cite{kirillov2023segment}. The Grounded-Segment-Anything (Grounded-SAM) model integrates object detection and segmentation by first detecting objects with Grounding DINO and then segmenting them using SAM, based on bounding boxes as textual prompts \cite{ren2024grounded}. MobileSAM adapts this framework by employing a smaller image encoder, TinyViT, and knowledge distillation, reducing the model size by 60 times while maintaining comparable performance to SAM \cite{zhang2023mobilesam}. 
Our approach, FastSAM, focuses on speeding up SAM's inference by transferring high-resolution features to low-resolution during clustering and reconstructing them back during inference, improving the process speed by up to 15 times.

\section{Methodology}
\begin{figure*}[t]  
  \centering  
  \includegraphics[width=\textwidth]{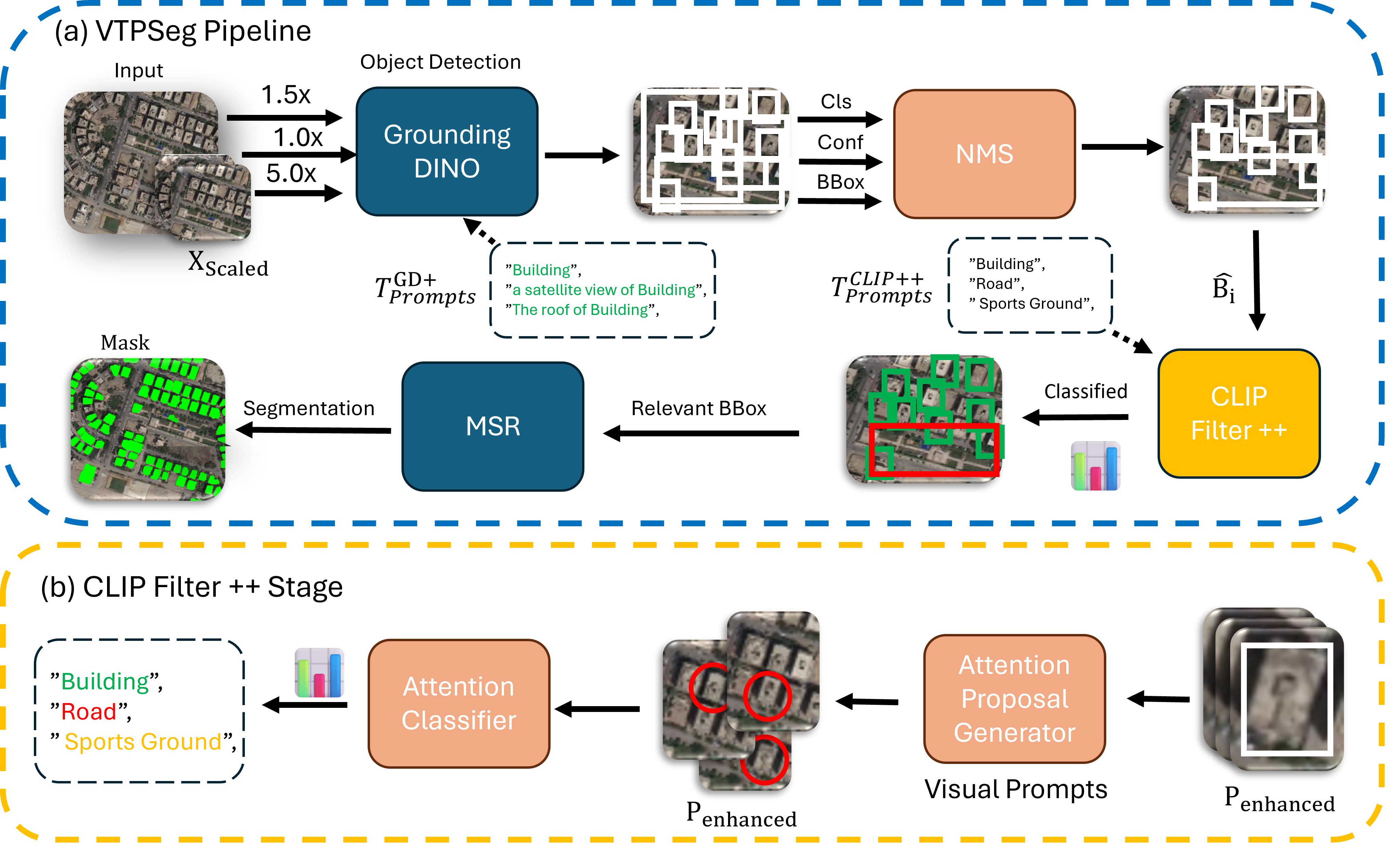}  
  \caption{(a) The overall pipeline of our VTPSeg, given a remote sensing image, the user needs to provide a set of segmented text queries of interest. Here, the text of interest is building. Multiple similar descriptions of the same target can effectively avoid detection misses, e.g., a roof of building. text prompts and multis-scale Patch of the image are fed into Grounding DINO for detection. Useless frames are then suppressed by the NMS method. This is followed by moving to the (b) CLIP Filter++ Stage, where a Visual Prompts module and an Attention Classifier module are included, the latter being used to evaluate the graphical alignment of each Visual Prompt.}
  \label{fig:myfig}
\end{figure*}

\subsection{The Proposed VTPSeg Framework}

The proposed VTPSeg architecture utilizes the zero-shot capabilities of GD+, CLIP Filter++, and FastSAM models to enhance segmentation in complex scenes, as shown in Fig. 2a. This pipeline processes the dataset as 
\[
X = \{x_i\}_{i=1}^N,
\]
where each image \( x_i \) is characterized by dimensions \( \mathbb{R}^{C \times H \times W} \) with channels \( C \), height \( H \), and width \( W \). The objective is to map these images to segmentation masks 
\[
Y = \{y_i\}_{i=1}^N,
\]
where each mask \( y_i \) encodes pixel-wise labels for various object categories.

Text prompts are defined as 
\[
T^{\text{VLMs}}_{\text{prompts}} = \{ \text{"desc 1"}, \text{"desc 2"}, \ldots, \text{"desc n"} \}
\]
where VLMs represent GD+ and CLIP++ models. These prompts describe object categories and are crucial for guiding the Grounding DINO+ and CLIP+ models, where "desc n" denotes the 
\( n^{\text{th}} \) object category. For Grounding DINO+, design as many synonyms as possible for each class; for CLIP++, the candidate words are designed for 
(1) task categories related prompts and 
(2) task categories related to remote sensing but unrelated to tasks' prompts.

Segmentation masks \( y_i \) are categorized within
\[
\{0, 1, 2, \ldots, \text{class number}\}^{1 \times H \times W}
\]
where "0" indicates background and subsequent integers signify specific objects such as buildings and cars. The "class number" depends on the diversity of objects within the dataset.

The Grounding DINO+ module, based on Grounding DINO, is used to solve three problems:
\begin{enumerate}
    \item High-resolution multi-scale recognition.
    \item Reducing missed detections by adopting extensive candidate words.
    \item Removing redundant and overly large candidate boxes after object detection.
\end{enumerate}
It is denoted as \( GD+(\cdot) \), which takes scaled images \( X_{\text{scaled}} \) and candidate prompts \( T^{\text{GD+}}_{\text{prompts}} \) as input:

\begin{equation}
\text{GD+}(\mathbf{X}_{\text{scaled}}, T^{\text{GD+}}_{\text{prompts}}) = \{\hat{C}_i, \hat{B}_i, \text{conf}_i \}
\end{equation}

The scaled images have scale factors of 0.5, 1.0, and 1.5. Smaller factors detect larger objects, while larger factors detect smaller objects. \( T^{\text{GD+}}_{\text{prompts}} \) are carefully designed by expanding the candidate semantic set through finding related descriptions or synonyms for each category in the task. 

For example, we can use the following synonyms or related statements to describe the "building" category:
\begin{equation} 
\{\text{"Roof"}, \text{"The roof of a building"}, \ldots, \text{"House"}\}.
\end{equation}

The predicted class \( \hat{C}_i \), bounding boxes (BBox) \( \hat{B}_i \), and confidence scores \( \text{conf}_i \) are the output pairs from the GD+ module. These include many more candidate boxes than actual targets, along with their confidence scores. We use non-maximum suppression (NMS) to preliminarily filter out redundant boxes.


The NMS method suppresses some candidate boxes but still cannot remove misclassifications or invalid boxes. To address these limitations, we propose an improved method based on CLIP, called CLIP Filter++, aimed at improving the accuracy of classification filtering. We use a combination of visual prompts and candidate word prompts instead of directly using the target category prompts of the CLIP model to filter the bounding boxes.

As shown in Figure 2b, first, to enable CLIP to capture contextual information and maintain global consistency, we use the Patch Extension method to expand the visual acquisition range of CLIP by a certain proportion. We define the patch extension results as the set of ${P^{{Ext}}_{i}}$. Then, by iteratively transforming the original BBoxes into red circles and adding them to each $P^{{Ext}}_{i}$, we guide the CLIP model's attention to individual objects, resulting in $x_{i}'$. We further explain the framework's operations through equations (2) to (5).

\begin{equation}
T^{\text{CLIP++}}_{\text{prompts}} = \{\text{Related categories, Unrelated categories}\}
\end{equation}

\setlength{\fboxrule}{0.7pt}
\noindent  
\fbox{%
  \begin{minipage}{3.3 in}

    \textbf{[Task Categories]}:\\
    \textit{\{Building, Road, Lake, Forest\}}
    \vspace{0.2cm}

    \textbf{[Candidate Prompts]}:\\
    \textit{\{Building, Road, Lake, Forest, Car, Cropland, Basketball court, Plain\}}
    \vspace{0.2cm}

    \textbf{[Explanation]}: \\
    \textit{The candidate prompts contain two parts: (1) task-related, and (2) not related to the task but related to the remote sensing area. This is to avoid CLIP selection bias due to the scarcity and inadequate description of categories. Therefore, the number of prompts will be much greater than the number of task categories. For further explanation, if the task is to classify "building" and "lake", we include prompts like "building", "ground" (unrelated), "lake", and "grass" (unrelated).}

  \end{minipage}%
}

\paragraph{CLIP Processing}
The enhanced patches are processed by CLIP as follows:
\begin{enumerate}
    \item Each $P_{\text{enhanced}}$, along with its red circle, is fed into the CLIP model. This visual modification aids CLIP in focusing its analysis on the object within the circle while still considering the contextual background.
    
    \item CLIP evaluates the similarity between the image content of each $P_{\text{enhanced}}$ and the set of text prompts $T^{\text{CLIP}}_{\text{prompts}}$, defined as:
    \[
        T^{\text{CLIP}}_{\text{prompts}} = \{\text{"The satellite view of \{class\_n\}"}\}.
    \]
    
    \item Internally, CLIP projects both the image embeddings of $P_{\text{enhanced}}$ and the embeddings of the text prompts into a shared embedding space. The similarity scores are then computed as the dot products between these embeddings. The closer the embeddings, the higher the similarity score, indicating a stronger match between the image and the text description.
    
    \item A softmax function is applied to these similarity scores to normalize them and determine the best matching text prompt. The index of the highest score is used to identify the best match, linking $P_{\text{enhanced}}$ to the corresponding category in $T^{\text{CLIP}}_{\text{prompts}}$.
\end{enumerate}



\subsection{Semantic Segmentation using FastSAM}


\begin{algorithm}
\caption{Semantic Segmentation with FastSAM (Pseudocode)}
\begin{algorithmic}
\State \textbf{Input:} List of bounding boxes \texttt{bboxes}, Image \texttt{image}
\State \textbf{Output:} List of segmentation masks \texttt{masks}
\State \texttt{masks} $\gets$ empty list
\For{each \texttt{bbox} in \texttt{bboxes}}
    \State \texttt{centroid} $\gets$ \texttt{calculate\_centroid(bbox)}
    \State \texttt{mask} $\gets$ \texttt{FastSAM(image, centroid)} \Comment{Apply FastSAM using the centroid as a point-prompt}
    \State \texttt{masks.append(mask)} \Comment{Store the generated mask}
\EndFor
\State \Return \texttt{masks}

\Procedure{calculate\_centroid}{\texttt{bbox}}
    \State $x_{\text{centroid}}$ $\gets$ (\texttt{bbox[0]} + \texttt{bbox[2]}) / 2
    \State $y_{\text{centroid}}$ $\gets$ (\texttt{bbox[1]} + \texttt{bbox[3]}) / 2
    \State \Return (\texttt{$x_{\text{centroid}}$, $y_{\text{centroid}}$})
\EndProcedure
\end{algorithmic}
\end{algorithm}

Once we have the final bounding box, we use FastSAM for the final segmentation. As described in Algorithm 1, we utilize the coordinates of the retained BBox to compute centroids for each BBox, which then serve as point-prompts for the FastSAM model. Since FastSAM can only handle one point-prompt at a time, we process these prompts sequentially in a loop. This method ensures precise and targeted segmentation of each identified object within the image, thereby enhancing the accuracy of our semantic segmentation.

\section{Experiments}
\begin{table*}[h]
\centering
\caption{Comparison of model performance in different remote sensing datasets (\%)}
\begin{tabular}{@{}lcccccc@{}}
\toprule
\multicolumn{6}{c}{\textbf{WHU Building Dataset}} \\
\midrule
Model & MIoU & Pixel Accuracy & Pixel Precision & Pixel Recall  & Dice Coefficient\\
\midrule
FastSAM(Everything-prompts) & \textbf{48.67} & 71.40 & 67.15  & \textbf{65.77} & \textbf{60.87} \\
MobileSAM(Anything-prompts) & 31.14 & 48.70 & 55.48  & 58.26 & 42.49  \\
Grounded-SAM & 31.71 & 54.06 & 60.43  & 54.27 & 42.84\\
\midrule
\textbf{VTPSeg(Ours)} & 47.83 & \textbf{72.15} & \textbf{68.34}  & 63.76 & 58.88 \\
\bottomrule
\multicolumn{6}{c}{\textbf{LoveDA Dataset}} \\
\midrule
Model & MIoU & Pixel Accuracy & Pixel Precision & Pixel Recall  & Dice Coefficient\\
\midrule
FastSAM(Everything-prompts) & 44.70 & \textbf{75.65} & 67.32 & 55.97 & 52.93 \\
MobileSAM(Anything-prompts) & 35.69 & 61.41 & 52.49  & 50.12 & 46.53  \\
Grounded-SAM & 44.70 & 75.64 & 67.32  & 55.97 & 52.93\\
\midrule
\textbf{VTPSeg(Ours)}  & \textbf{53.22} & 75.37 & \textbf{68.01}  & \textbf{71.91} & \textbf{64.82} \\
\bottomrule
\multicolumn{6}{c}{\textbf{Inria Aerial Image Labeling Dataset}} \\
\midrule
Model & MIoU & Pixel Accuracy & Pixel Precision & Pixel Recall  & Dice Coefficient\\
\midrule
FastSAM(Everything-prompts) & 37.70 & \textbf{66.65} & 61.47 & 53.72 & 46.88 \\
MobileSAM(Anything-prompts) & 35.69 & 61.41 & 52.49 & 50.12 &  46.53 \\
Grounded-SAM & 37.70 & 61.47 & 60.43  & 53.72 & 46.88\\
\midrule
\textbf{VTPSeg(Ours)}  & \textbf{46.62} & 56.19 & \textbf{66.77}  & \textbf{58.87} & \textbf{56.99} \\
\bottomrule
\multicolumn{6}{c}{\textbf{xBD Dataset}} \\
\midrule
Model & MIoU & Pixel Accuracy & Pixel Precision & Pixel Recall  & Dice Coefficient\\
\midrule
FastSAM(Everything-prompts) & 40.80 & 71.35 & 55.59  & 62.91 & 48.08 \\
MobileSAM(Anything-prompts) & 31.14 & 48.70 & 55.48  & 58.26 & 42.49  \\
Grounded-SAM & 36.90 & 65.55 &  65.98 & 43.80 & 41.76 \\
\midrule
\textbf{VTPSeg(Ours)}  & \textbf{50.19} & \textbf{90.63} & \textbf{66.17} & \textbf{67.02} & \textbf{62.49} \\
\bottomrule
\end{tabular}
\end{table*}
\subsection{Experimental Setting}
\paragraph{Datasets.} 
In this paper, we do this in a common semantic segmentation dataset of remotely sensed building. Which are WHU Building \cite{WHU}, LoveDA \cite{LoveDA}, Inria Aerial Image \cite{AerialDataset}, xBD \cite{xBD} and iSAID\cite{zamir2019isaid}.

The WHU Building Dataset specifically utilizes the "Satellite dataset I (global cities)" for experiments, comprising 204 satellite images of buildings, each sized 512x512 pixels. This dataset showcases a diversity in atmospheric conditions, multispectral fusion algorithms, and seasonal variations, which are instrumental in evaluating the robustness of building segmentation algorithms.

The LoveDA Dataset encompasses 5987 high spatial resolution (0.3 meters) remote sensing images captured over Nanjing, Changzhou, and Wuhan in China. It features two subsets: rural and urban, each containing multi-scale objects against complex backgrounds, offering a comprehensive challenge for segmentation tasks.

The Inria Aerial Image Labeling Dataset contains 360 TIF images with dimensions of 5000x5000, split equally into training and test sets. Covering various urban residential areas—from densely populated regions to mountainous towns—these images present varied lighting conditions, urban landscapes, and seasonal settings throughout the year.

The xBD Dataset includes over 850,000 building polygons from six different types of natural disasters worldwide, providing paired images (pre- and post-disaster) for analysis. For the purposes of this study, we solely utilize the pre-disaster building images from the test set, which consists of 1866 remote sensing images, each 1024x1024 in size. This subset is crucial for developing algorithms capable of accurate building detection and segmentation under crisis scenarios.

The iSAID is a large-scale instance segmentation benchmark dataset adapted from the DOTA\cite{DOAT} dataset, containing 15 categories, 2806 high-resolution images, and 655,451 labeled instances.

\subsection{Implementation details.} 
In our experiments, Grounding DINO utilized the Swin-B model, while CLIP employed the ViT-L-14-336 model. FastSAM used the FastSAM-x weights. We applied multi-scale transformations with scales of 0.5 and 1.0. The NMS parameters were set with an overlap threshold of 0.1. Patch Enhancement was applied at magnification factors of 1.2 and 1.5. All experiments were conducted on NVIDIA 4090 GPU.


\subsection{Metrics Comparison}
We evaluated our proposed VTPSeg pipeline on remote sensing datasets pertaining to urban buildings, rural landscapes, and disaster assessment, specifically on the WHU Building Dataset, LoveDA Dataset, Inria Aerial Image Labeling Dataset, and xBD Dataset. In all these datasets, VTPSeg excelled across all evaluation metrics, surpassing other existing zero-shot methods.
As shown in Table 1, In the relatively simpler xBD Dataset, VTPSeg's Pixel Accuracy showed a significant lead, with a difference of 19.28\% and 41.93\% over FastSAM and the poorest performer MobileSAM, respectively. Additionally, its Dice Coefficient also demonstrated a remarkable advantage.
On the more challenging LoveDA Dataset, VTPSeg continued to perform excellently, achieving an MIoU of 53.22\%, which is 8.52\% higher than FastSAM (Everything-prompts); the Dice Coefficient increased from 52.93\% to 64.82\%, showcasing our method's effectiveness in handling this dataset. Moreover, in the Inria Aerial Image Labeling Dataset and xBD Dataset, VTPSeg also showed overall performance enhancements, with MIoU improvements of 8.92\% and 9.39\%, respectively, proving the efficacy and generalizability of our VTPSeg.

\subsection{Visualization comparisons}
\begin{figure*}[t]  
  \centering  
  \includegraphics[width=\textwidth]{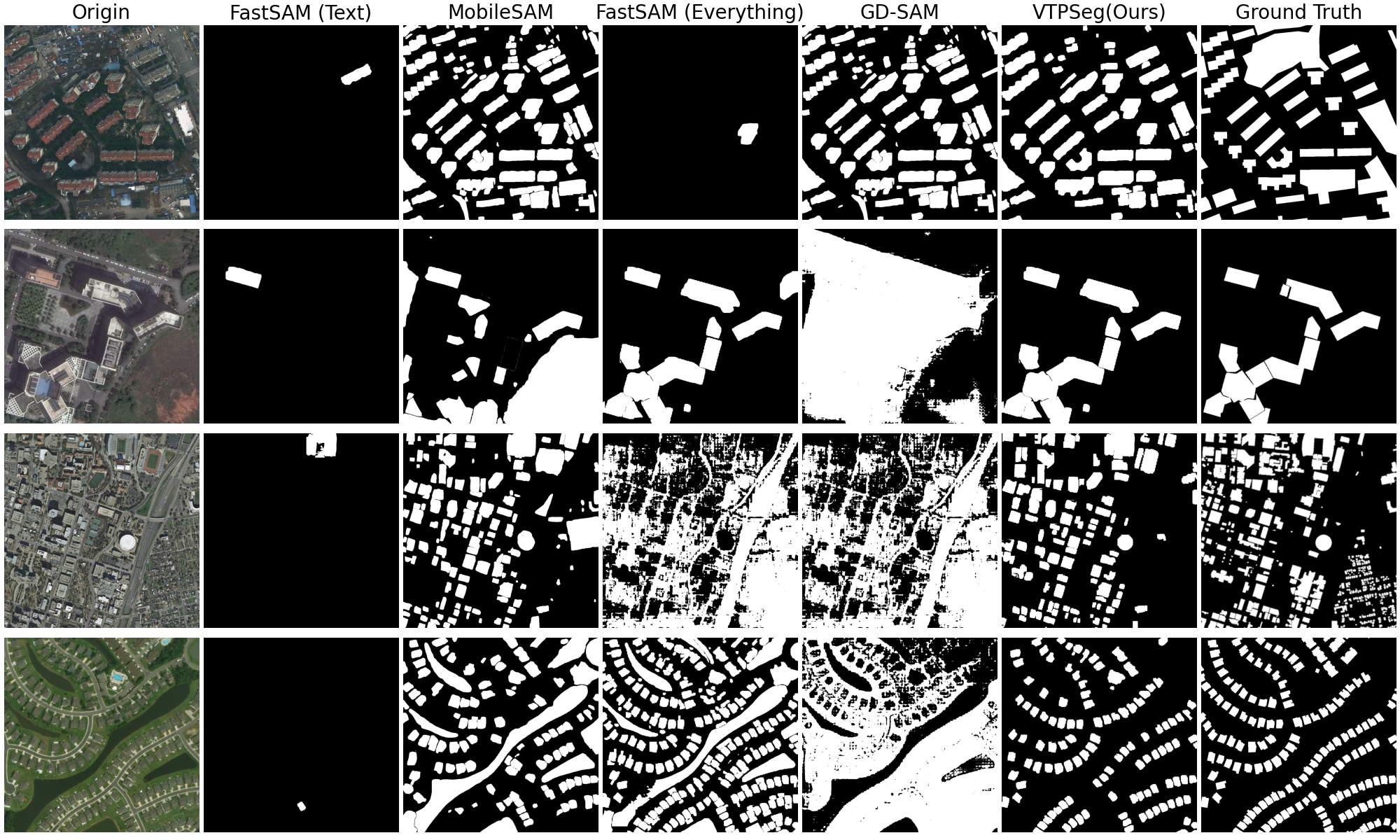}  %
  \caption{Comparison of visualization results from FastSAM (Everything and text prompts), MobileSAM, Grounded-SAM, and VTPSeg across four datasets. It is clear that our VTPSeg consistently delivers significantly outcomes.}
  \label{fig:myfig}
\end{figure*}
We present the visualization results for various state-of-the-art models in the field of zero-shot remote sensing, further demonstrating the efficacy of our proposed method. In Fig. 3., it illustrates that the FastSAM model, which allows for language prompts, can only segment out one object deemed to have the highest image-text similarity per input. The results from FastSAM and MobileSAM for segmenting 'Everything' and 'Anything' are competitive, but the lack of refined categorization leads to incorrect classifications of squares and grasslands. The Grounded-SAM model, which utilizes language-guided multi-object segmentation, performs unsatisfactorily with small-scale targets. In contrast, our proposed VTPSeg not only possesses the capability to segment 'Everything', but also can filter masks based on user language prompts.

\section{Ablation experiments}
\begin{table}[h]
\centering
\caption{Comparison of Incremental Component Performances (\%)}
\begin{tabular}{@{}l@{\hspace{4pt}}c@{\hspace{5pt}}c@{\hspace{5pt}}c@{\hspace{5pt}}c@{\hspace{5pt}}c@{\hspace{4pt}}@{}}
\toprule
Model                              & MIoU  & PA & PP & PR & Dice \\
\midrule
GD+FastSAM                         & 48.9 & 76.9 & 67.2 & 61.9 & 58.6            \\
GD+(Scal+NMS)+FastSAM              & 49.4 & 71.7 & 64.7 & 70.1 & 61.7          \\
GD+(Scal+NMS)+CLIP Filter++ + FastSAM & 51.1 & 73.2 & 66.0 & 69.73 & 62.64          \\
\textbf{VTPSeg}                    & \textbf{53.2} & \textbf{75.4} & \textbf{68.0} & \textbf{71.9} & \textbf{64.8} \\
\bottomrule
\end{tabular}
\end{table}

In this section, we conduct an ablation study to evaluate the contribution of 4 components in our semantic segmentation models. 
As shown on Table 2. The results indicate progressive improvements across all metrics with the integration of each component. The addition of scaling and NMS notably enhanced the pixel recall from 61.9\% to 70.1\%, suggesting better recovery of relevant features. The introduction of CLIP Filter++ increased the MIoU significantly by 1.7\%, indicating a refined accuracy in predicting the segment boundaries.

The VTPSeg model, which integrates all the tested components, achieved the highest scores in all evaluated metrics, demonstrating the effectiveness of combining these techniques for semantic segmentation. The model not only improved MIoU by 4.3 points compared to the baseline but also showed a balanced improvement across precision and recall, confirming its robustness in various segmentation scenarios.

\section{More experiments}
\begin{table*}[h]
\centering
\caption{Comparison of model performance in iSAID datasets (\%)}
\begin{tabular}{@{}l|cccccccccccccccc@{}}
\toprule
\textbf{Model} & \textbf{MIoU} & \textbf{SP} & \textbf{ST} & \textbf{BD} & \textbf{TC} & \textbf{BC} & \textbf{GF} & \textbf{BR} & \textbf{LV} & \textbf{SV} & \textbf{HC} & \textbf{SP} & \textbf{RA} & \textbf{SF} & \textbf{PL} & \textbf{HR} \\
\midrule
FCN-8s \cite{fcn} & 38 & 51 & 23 & 26 & 75 & 30 & 28 & 8 & 49 & 37 & 0 & 31 & 52 & 52 & 63 & 42 \\
SPGNet \cite{spgnet} & 43 & 53 & 43 & 59 & 75 & 49 & 44 & 11 & 53 & 31 & 4 & 40 & 34 & 60 & 45 & 46 \\
Deeplab v3 \cite{deeplab} & 56 & 60 & 50 & 77 & 84 & 58 & 59 & 33 & 55 & 34 & 31 & 45 & 66 & 72 & 76 & 45 \\
U-Net \cite{u-net} & 58 & \textbf{64} & 53 & 67 & \textbf{87} & 57 & 50 & 34 & \textbf{59} & \textbf{48} & 30 & 42 & 70 & 70 & \textbf{82} & \textbf{54} \\
\midrule
VTPSeg(RemoteCLIP) & 55 & 33 & 43 & \textbf{83} & 77 & 66 & 77 & 1 & 37 & 26 & 22 & 68 & 94 & 71 & 82 & 40 \\
VTPSeg(Ours) & \textbf{61} & 26 & \textbf{54} & 76 & 77 & \textbf{69} & \textbf{79} & \textbf{62} & 44 & 31 & \textbf{22} & \textbf{76} & \textbf{97} & \textbf{80} & 81 & 45 \\
\bottomrule
\end{tabular}
\end{table*}
To further explore the applicability of our method in more complex scenarios, We evaluate our approach on the complex iSAID dataset, which encompasses 15 diverse classes of remotely sensed objects. This dataset is particularly challenging due to the variability and intricacy of the object classes it contains.  We benchmark our model against several well-established semantic segmentation frameworks such as FCN-8s, SPGNet, Deeplab v3, and U-Net. 
Table 3 presents a detailed comparison of model performance on the iSAID dataset. Our method is more balanced in all category segmentation, and there is no situation where segmentation is not possible or the segmentation effect is very poor.
Although models like RemoteCLIP\cite{remoteclip} are pre-trained on remote sensing images, they can struggle with accurate recognition in scenes with multiple targets. To address this, we enhanced the CLIP model with patch enhancements, adding more contextual information. In tests on the iSAID Val Set with the same model size (ViT-L-14-336), CLIP outperformed RemoteCLIP by 6\%.

\section{Conclusion}
In this paper, we introduced VTPSeg, a zero-shot multi-modal segmentation framework for remote sensing. By leveraging the strengths of Grounding DINO+, CLIP Filter++, and FastSAM, VTPSeg addresses challenges like multi-scale object detection, attention drift, and mask redundancy. Ablation studies demonstrate the effectiveness of key component contributions. Experiments on multiple datasets show that VTPSeg outperforms existing methods in MIoU and pixel accuracy. Future work will focus on optimizing VTPSeg for large-scale tasks and integrating data from diverse remote sensing sources to improve generalization.

%








\bibliographystyle{splncs04}
\bibliography{mybibliography}
\end{document}